\documentstyle[aps]{revtex}
\def \be   {\begin{equation}}
\def \ee   {\end{equation}}
\def \l {\label}
\begin{document}
\input epsf
%\hskip7.5cm {UFES-DF-MMS.97/3}\\
\baselineskip=20pt
%\draft
%\preprint
\title{Gauge fields in a discrete approach}
\author{Manoelito M de Souza\footnote{Supported by CAPES/COPLAG-ES}}
\address{Universidade Federal do Esp\'{\i}rito Santo - Departamento de
F\'{\i}sica\\29065.900 -Vit\'oria-ES-Brasil}
\date{\today}
\maketitle
\begin{abstract}
\noindent A discrete field formalism exposes the physical meaning and origins of gauge fields, their symmetries and singularities. They represent a lack of a stricter field-source coherence.
\end{abstract} 
\begin{center}
\end{center}
We want to discuss here the physical meaning and the origins of gauge field symmetries and singularities.
It is well known that gauge freedom seems to be a consequence of the field anti-symmetry but what makes anti-symmetric all fundamental interaction fields? In General Relativity the potential, but not the field is described by a symmetric tensor. In this work we discuss the physical interpretation of a discrete point-like field introduced through a new integral transform of the potential field in order to show that its gauge symmetry and singularity are  consequences of causality and Lorentz invariance and that they reflect a loss of field-source coherence. The discrete fields are proposed to be associated to the field quanta as they are their closest description possible in a classical theory. With discrete fields field theory loses its ubiquitous problems with infinities.
\begin{center}
Causality: local and extended
\end{center}
\noindent Any given pair of events on Minkowski spacetime defines a four-vector $\Delta x.$ If this  $\Delta x$ is connected to the propagation of a free physical object (a signal, a particle, a field, etc) it is constrained to satisfy
\be
\label{1}
\Delta\tau^2=-\Delta x^{2}, 
\ee 
where $\tau$ is a real-valued parameter. We use the metric $\eta=diag(1,1,1,-1)$. So, (\ref{1}) just expresses that $\Delta x$ cannot be spacelike. A physical object does not propagate over a spacelike $\Delta x.$ This is {\it local causality}.  Geometrically it is the definition of a three-dimensional double cone; $\Delta x$ is the four-vector separation between a generic event $x^{\mu}\equiv({\vec x},t)$ and the cone vertex. This conic hypersurface, in field theory, is the free-field support: a {\bf free} field cannot be inside nor outside but only on the cone. The cone-aperture angle $\theta$ is given by $\tan\theta={|\Delta {\vec x}|}/{|\Delta t|},\; c=1,$ or equivalently, $\Delta\tau^{2}=(\Delta t)^{2}(1-\tan^{2}\theta).$ A change of the supporting cone corresponds to a change of speed of propagation and is an indication of interaction.
Special Relativity restricts $\theta$ to the range $0\le\theta\le{\pi}/{4},$ which corresponds to a restriction on $\Delta\tau:$ $0\le|\Delta\tau|\le|\Delta t|.$ The lightcone ($\theta={\pi}/{4},$ or $|\Delta\tau|=0$) and the t-axis in the object rest-frame ($\theta=0,$ or $|\Delta\tau|=|\Delta t|$) are the extremal cases. We want to work however with a more restrictive constraint:
\be
\l{f}
\Delta\tau+  f.\Delta x=0,
\ee
where $f$ is defined by $f^{\mu}={dx^{\mu}}/{d\tau},$ a constant  four-vector tangent to the cone; it is  timelike $(f^{2}=-1$) if $\Delta\tau\ne0,$  or lightlike $(f^{2}=0$ in the limit) if $\Delta\tau=0$. \\
The equation (\ref{f}) defines a hyperplane tangent to the cone (\ref{1}). Together, (\ref{1}) and (\ref{f}) define a cone generator $f$, tangent to $f^{\mu}$. A fixed four-vector $f^{\mu}$ at a point represents a fibre in the spacetime, a straight line tangent to $f^{\mu}$, the $f$-generator of the local cone (\ref{1}).\\
Extended causality is the imposition of both (\ref{1}) and (\ref{f}) to the propagation of any point-like physical object. Geometrically, it is a requirement that the object remains on the cone generator $f$.\\ 
\noindent The lightcone is the support for the propagation of a massless field; the field itself and its theory are defined over the lightcone. With extended causality we have to consider not the lightcone but its generators and to define a field with support on a generic fibre $f$, a $(1+1)$-manifold embedded on a $(3+1)$-Minkowski spacetime. It is a discrete field $A_{f}(x,\tau)$ which is related to the standard continuous field $A(x,\tau)$ through an integral transform to be introduced later.
As a consequence of the causality constraint (\ref{f}), the fields must be explicit functions of x and of $\tau,$ where $\tau$, is a supposedly known function of x, a solution of (\ref{1}):  $\tau=\tau_{0}\pm\sqrt{-(\Delta x)^{2}}.$ We have from (\ref{f}) that 
\be
\l{fmu}
f_{\mu}=-\frac{\partial\tau}{\partial x^{\mu}}.
\ee
For a massless field, as it propagates without a change on its proper time, $\Delta\tau=0$, $\;\tau$ is actually the instantaneous proper-time of its source at the event of its emission.  Well-known examples of this are the Li\`enard-Wiechert solutions\cite{Rorhlich,Jackson,Teitelboim,Rowe,Lozada}, discussed in this context in \cite{hep-th/9610028}.
\begin{center}
Causality and dynamics
\end{center}
The constraint (\ref{f}) has a very important dynamical content as we discuss now. For a massless field emitted by a point charge on a worldline $z(\tau)$, parameterized by its proper-time $\tau$, $\Delta\tau=0$ and $\Delta x=x-z(\tau).$ The restriction (\ref{f}) is then reduced to $f.(x-z(\tau))=0$ and this implies that the event $x$, where the field is being observed, and the charge retarded position  $z(\tau)$ must belong to a same null line $f$. It is not necessary to explicitly distinguish a generic $\tau$ from a $\tau$ at a retarded position, as the situations considered in this note always refer to the last one.
More information can be extracted from this constraint as $\partial_{\mu}f.(x-z){\big |}_{f}=0$ implies, with (\ref{fmu}), on 
\be
\l{fv} 
f.V{\big |}_{f}=-1,
\ee
where $V={dz}/{d\tau}$. This relation  may be seen as a covariant normalization to 1 of the time component of $f$ in the source rest-frame at its retarded time, 
$
f^{4}{\big |}_{{f\;\;}\atop{{{\vec V}=0}}}=|{\vec f}|{\big |}_{{f\;\;}\atop{{\vec V}=0}}=1.
$
With a further derivation and with $a^{\mu}={dV^{\mu}}/{d\tau}$ we get from (\ref{fv}) the following most important relationship
\be
\l{dA0}
a.f{\big |}_{f}=0,
\ee
between the direction $f$ along which the signal is emitted (absorbed) and the instantaneous change in the charge state of motion at the retarded (advanced) time. (\ref{dA0}) is a causal constraint on the propagation of any field, regardless its tensorial nature, which we did not specify yet. It implies that
$
a_{4}=({{\vec a}.{\vec f}}/{f_{4}}){\big|}_{f},
$
whereas $a.V\equiv0$ leads to
$a_{4}=({{\vec a}.{\vec V}}/{V_{4}}){\big|}_{f},$
and so we have that in the charge instantaneous rest frame at the emission time ${\vec a}$ and ${\vec f}$ are orthogonal vectors,
$
{\vec a}.{\vec f}{\big |}_{{f\;\;}\atop{{\vec V}=0}}=0.
$
\begin{center}
Field equations
\end{center}
 \noindent (\ref{1}) implies that $\tau$ is a function of $x$ whereas (\ref{f}) implies on (\ref{fmu}). Then the derivatives of $A_{f}(x,\tau),$ allowed by the constraint (\ref{f}), are the directional derivatives along $f,$ which with the use of (\ref{fmu}) we write as
$$
\partial_{\mu}A(x,\tau){\big |} _{f}=(\frac{\partial }{\partial x^{\mu}}+\frac{\partial \tau}{\partial x^{\mu}}\frac{\partial}{\partial \tau})A(x,\tau){\big |} _{f}={\big(}\frac{\partial }{\partial x^{\mu}}-f_{\mu}\frac{\partial}{\partial \tau}{\big)}A_{f}\equiv\nabla_{\mu} A_{f}.
$$
With $\nabla$ replacing $\partial$ for taking care of the constraint (\ref{f}), the propertime $\tau$ can be treated as a fifth independent  coordinate. We adopt this geometrical approach. The field equation for a massless discrete field is
\be
\label{wef}
\eta^{\mu\nu}\nabla_{\mu}\nabla_{\nu}A_{f}(x,\tau)=J(x,\tau),
\ee
or, explicitly
$(\eta^{\mu\nu}\partial_{\mu}\partial_{\nu}-2f^{\mu}\partial_{\mu})A_{f}=J,$
as $f^{2}=0$. $J$ is its source. The $f$-wave equation (\ref{wef}) can be solved by an f-Green's function,
$
A_{f}(x,\tau_{x})=\int d^{4}yd\tau_{y}\; G_{f}(x-y,\tau_{x}-\tau_{y})\;J(y),
$
where the sub-indices specify the respective events $x$ and $y$, and $G_{f}(x-y,\tau_{x}-\tau_{y})$ being a solution of
$
\eta^{\mu\nu}\nabla_{\mu}\nabla_{\nu}G_{f}(x-y,\tau_{x}-\tau_{y})=\delta^{4}(x-y)\delta(\tau_{x}-\tau_{y}):=\delta^{5}(x-y).
$
This equation has \cite{ecwpd}:
\be
\l{pr9}
G_{f}(x,\tau)=\frac{1}{2}\theta(-b{\bar f}.x)\theta(b\tau)\delta(\tau+  f.x)=\frac{1}{2}\theta(bf^{4}t)\theta(b\tau)\delta(\tau+  f.x),
\ee
as a solution, 
where $b =\pm1,$ and $\theta (x)$ is the Heaviside function, $\theta(x\ge0)=1$ and $\theta(x<0)=0.$  $G_{f}(x,\tau)$ does not depend on ${\vec x}_{\hbox {\tiny T}}$, where the subscript ${\hbox {\tiny T}}$ stands for transversity with respect to ${\vec f}$: ${\vec f}.{\vec x}_{{\hbox {\tiny T}}}=0.$  This justifies this approach of working with discrete field; its propagation does not depend on ${\vec x}_{\hbox {\tiny T}}$, or more explicitly, on the parts of spacetime that are not in $f$. Two distinct fields emitted by two neighbouring point-sources do not see each other; each one of them can be treated as an independent single entity. It implies also, as it will be shown later, that the energy-momentum content of $A_{f}$ is everywhere conserved. This strongly suggests seeing $A_{f}$ as the fundamental field, a classical description of the quanta of the field $A$.\\
For $f^{\mu}=({\vec f}, f^{4})$, ${\bar f}$ is defined by ${\bar f}^{\mu}=(-{\vec f}, f^{4});$ $f$ and ${\bar f}$ are two opposing generators of a same lightcone; they are associated, respectively, to the $b=+1$ and to the $b=-1$ solutions and, therefore, to the processes of creation and annihilation of a discrete field $A_{f}$. $\theta(-b{\bar f}.x)=\theta(bt).$
For $b=+1$ or $t>0$, $G_{f}(x,\tau)$ describes a point signal emitted by the charge  at $\tau_{ret}=0,$ and that has propagated to $x$ along the fibre $f,$ of the future lightcone of $z(\tau_{ret})$;  for $b=-1$ or $t<0,$  $G_{f}(x,\tau)$ describes a point signal that is propagating  along the fibre $\bar{f}$ of the future lightcone of $x$ towards the point $z(\tau_{adv})$ where it will be absorbed (annihilated) by the charge. 
Observe the differences from the standard interpretation of the Li\`enard-Wiechert solutions with local causality. There is no advanced, causality violating solution here. $J$ is the source of the $f$ solution and a sink for the $\bar{f}$ solution. These two solutions correspond to creation and annihilation of $A_{f}$. $G_{f}$ and its properties will be discussed in more details in \cite{ecwpd}.
\begin{center}
Discrete fields
\end{center}
In a generic way an extended source is a set, continuous or discrete, of point sources and, according to (\ref{pr9}), each one of them may generate its independent $A_{f}$ field. But the essence of this discrete formalism is that fields and sources are discrete, point-like; this might be taken as an extra assumption but actually any extended object, as far as our technology can tell, is made of point-like objects. The proton structure is a good illustration of what we mean; it is an extended object but actually made of points: quarks and gluons. The image of a continuous extended object is just a macroscopic approximation, an average output of our senses or measuring apparatus. Thus let us now apply this $f-$formalism to the field generated by a  point scalar charge. Just for the sake of simplicity we will fix $A$ and $A_{f}$ as vector fields like the four-vector potential in Maxwell theory, but we will take eq. (\ref{wef}) as our departure point for studying the field, assuming that we don't know anything else about it.\\ With $\tau$ being treated as an independent fifth parameter, a definition of a four-vector current must carry an additional constraint expressing the causal relationship between two events $y$ and $z$. Its four-vector current is given by
$J^{\mu}(y,\tau_{y}=\tau_{z})= eV^{\mu}(\tau_{z})\delta^{3}({\vec y}-{\vec z})\delta(t_{y}-t_{z}),$
where $z^{\mu}(\tau_{z}),$ is the electron worldline parameterized by its proper-time $\tau_{z}.$ In this definition of $J$, $\tau_{y}$ has to be equal to $\tau_{z}$ as a consequence of the Dirac deltas and of (\ref{1}). For $b=+1$, that is, for the field emitted by J we have
$
\l{Af}
A_{f}(x,\tau_{x})=2e\int d^{5}y G_{f}(x-y)V^{\mu}(\tau_{y})\delta^{3}({\vec x}-{\vec y})\delta(t_{x}-t_{y}),
$
where the factor 2 accounts for a change of normalization as we are now excluding the annihilated $A_{f}$ (the future lightcone). Then,
\be
\l{Af1}
A_{f}(x,\tau_{x})=eV^{\mu}(\tau_{z})\theta(t_{x}-t_{z})\theta(\tau_{x}-\tau_{z}){\big |}_{{\tau_{z}=\tau_{x}+f.(x-z)}}.
\ee
To the standard continuous field $A(x,\tau)$ one can add arbitrary solution from its homogeneous equation; this is compatible with its gauge freedom and it may indeed be necessary for attending some specific boundary conditions. For $A_{f}$ this is not so; it describes interactions between point objects for which boundary conditions have no meaning. This is not possible for $A_{f}$ because the solution to its homogeneous equation is just $\delta(\tau+f.x),$ which is trivial in the sense that $\nabla_{\mu}(\tau+f.x)\equiv0$. Besides it is a distribution, in contradistinction to (\ref{Af1}) which is a finite point-like field; so it cannot be used to redefine $A_{f}$. This ``constant" solution should not be added to (\ref{dA0}) in anyway because it would artificially introduce infinities where there is none and could destroy the theory consistency.
So, the field $A_{f}$ is given, essentially, by the charge times its four-velocity at its retarded time. For a massless field $\tau_{x}=\tau_{z}$ and $f.(x-z)=0.$  $\nabla\theta(t)$ and $\nabla\theta(\tau)$ do not contribute to $\nabla A_{f},$ except at $x=z(\tau),$ as a further consequence of the field constraints \cite{ecwpd}. Thus for $\tau=0$ and $t>0$ we write just 
$\nabla_{\nu}A^{\mu}_{f}=\nabla_{\nu}(eV^{\mu}){\big |}_{f}=-ef_{\nu}a^{\mu}{\big |}_{f}.$
Therefore $A_{f}$ is a divergenceless field, 
\be
\l{gc}
\nabla.A_{f}=0,
\ee
as a direct consequence of (\ref{dA0}). This is a very important point and deserves further elaboration. There is a causal link, a coherence, between $A_{f}$ and its source that leads necessarily to (\ref{gc}). This link does not depend on the field tensorial nature.  The extended causality constraint, i.e. the explicit constraints on a causal propagation of a point object, whichever be it, leads to the constraint (\ref{dA0}) between its direction of propagation $f$ and the change in the state of motion of its source (sink) at the emission (absorption) time. For the same reason $\nabla.J=0.$ We have charge conservation regardless the Maxwell tensor antisymmetry which supposedly we still don't know. Therefore charge conservation is also a consequence of (extended) causality and not of gauge symmetry, which we do not have yet.\\
 $A_{f}$ of (\ref{Af1}) is just an expression of the ``charge state of motion" at the emission time.  ``State of motion" is a relative or frame dependent concept and this suits well with $A_{f}$ being a potential; the proper field, the force carrier, is then associated to the charge acceleration, an absolute concept. We have started from eq. (\ref{wef}) assuming that we don't know anything about the field except that it is massless and must somehow be associated to the gradient of $A_{f}$ and, therefore, to the components of a second-rank (in the case of $A_{f}$ being a vector field) tensor 
$F^{\mu\nu}_{f}\Leftrightarrow-\nabla^{\nu}A_{f}^{\mu}=ef^{\nu}a^{\mu}{\big |}_{f},$
defined by two four-vectors, the acceleration of its source at its emission time and $f$. Therefore, $F_{f}$ could in principle be a scalar or either a symmetric or an antisymmetric tensor. But $a$ and $f$ are not independent as they are constrained by (\ref{dA0}), which besides excluding the scalar component  requires that ${\vec a}.{\vec f}=0,$ in the charge instantaneous rest-frame and this, as we show now, excludes the symmetric component. In other words, $F_{f}$ is a function not of $a$ but of $a_{\hbox{\tiny T}}$:
\be
\l{ft}
F_{f}{\big |}_{{\vec V}=0}
\Leftrightarrow efa_{\hbox{\tiny T}}{\big |}_{f}=ef(a-f\frac{{\vec a}.{\vec f}}{f_{4}^2}){\big |}_{f},
\ee
 as $a_{\hbox{\tiny T}}:=a-f\frac{{\vec a}.{\vec f}}{f_{4}^2}$. In this particular  frame the direction of propagation of $A_{f}$ is perpendicular to the electron instantaneous acceleration;  $A_{f}$ is a transversal field. But $ff\frac{{\vec a}.{\vec f}}{f_{4}^2}$ is not Lorentz  covariant (neither $a_{\hbox{\tiny T}}$) and there should be no privileged frame;  so it should not appear in the $F_{f}$ definition. Lorentz covariance requires an anti-symmetric $F^{f}_{\mu\nu}:=-e(f_{\mu}a_{\nu}-f_{\nu}a_{\mu}){\big |}_{f}$. Thus, 
\be
\l{me}
F^{f}_{\mu\nu}:=\nabla_{\mu}A^{f}_{\nu}-\nabla_{\nu}A^{f}_{\mu}.
\ee
That $F_{f}$ must be an antisymmetric tensor is a direct consequence of Lorentz covariance and of (\ref{dA0});  it cannot be proved in a context of local causality.\\
Observe that initially with just the field equation (\ref{wef}) we had no ground to talk about gauge freedom but even now after knowing that $A_{f}$ is just a potential and that the physical field $F_{f}$ must be antisymmetric there is still no gauge freedom, not even a residual one. 
$A_{f}$, despite the antisymmetry of $F_{f}$, is determined by the state of motion of its source. The link between $A_{f}$ and the state of motion of its source (a single point charge) does not allow any non-trivial gauge freedom. Extended causality incorporates into the background geometry the constraints that, in an explicitly covariant way, eliminate the field spurious degrees of freedom.\\
Having $F_{f}$ we  can get its energy-momentum tensor. After (\ref{me}) and (\ref{gc}) it is given by  $\Theta_{f}^{\mu\nu}=F_{f}^{\mu\alpha}F_{f}^{\beta\nu}\eta_{\alpha\beta}-\frac{\eta^{\mu\nu}}{4}F_{f}^{\alpha\beta}F^{f}_{\alpha\beta}=-e^{2}f^{\mu}f^{\nu}a^{2}|_{f}$ and satisfies $\nabla_{\nu}\Theta_{f}^{\mu\nu}=0$. The energy-momentum content of $A_{f}$ is everywhere conserved; $\Theta_{f}$ is finite and represents a point object changelessly propagating along $f$: a classical photon.
All that we have used, besides the wave equation (\ref{wef}), is causality and Lorentz covariance. Therefore we conclude that causality and Lorentz covariance  leads to, let's say, ``Maxwell's theory on a fibre $f$" with a fixed Lorentz gauge. 
\begin{center}
From discrete to continuous fields
\end{center}
\noindent Now we discuss the connection between this ``Maxwell's theory on $f$" and the Maxwell's theory, between the field $F_{f}$ and the $F$ that we measure in a laboratory. We can say, figuratively, that the lightcone is the union of all its generators. In a similar way the standard continuous field $A(x,\tau)$ represents the collection of all $A(x,\tau)_{f}$ from all possible fibres $f.$
 In such a picture  $A_{f}$ is the intersection of $A(x,\tau)$  with the fibre $f:$ 
\be
\label{Aff}
A(x,\tau)_{f}=A(x,\tau){\big |}_{f},
\ee
it is the restriction of $A(x,\tau)$ to f. It represents an element of $A(x,\tau)$, the part of it contained in the fibre $f$: a point propagating along $f.$ The converse of (\ref{Aff}) is given by
\be
\label{s1s}
A(x,\tau)=\frac{1}{2\pi}\int d^{4}f\;\delta(f^{2})A(x,\tau)_{f},
\ee
where 
$
d^{4}f=df_{4}\;|{\vec f}|^{2}\;d|{\vec f}|\;d^{2}\Omega_{f}
$
and 
$
\delta(f^{2})\equiv1/({2|{\vec f}|})\;\{\delta(f^{4}-|{\vec f}|)+\delta(f^{4}+|{\vec f}|)\}.
$
For the emitted $(f^{4}=|{\vec f}|)$ field in the source instantaneous rest frame at the emission time $(f^{4}=1)$ the equation (\ref{s1s}) can be written as
\be
\label{s}
A(x,\tau)=\frac{1}{4\pi}\int d^{2}\Omega_{f}A(x,\tau)_{f},
\ee
where the integral represents the sum over all directions of ${\vec f}$ on  a lightcone. $4\pi$, we see, is a normalization factor and (\ref{s}) is a particular case of
$
A(x,\tau)=({\int_{\Omega_{f}} d^{2}\Omega_{f}A(x,\tau)_{f}})/({\int_{\Omega_{f}} d^{2}\Omega_{f}}).$ (Operationally $\Omega_{f}$ is  defined by the asperture of our measuring apparatus.)
On the other hand an integration over the $f$ degrees of freedom in (\ref{wef}) and (\ref{me}) with the use of (\ref{s}) reproduces, respectively, the usual wave equation and the Maxwell field of the standard formalism, 
 as $\int d^{2}\Omega_{f}f^{\mu}\partial_{\mu}\partial_{\tau}A_{f(x)}=0$ because $A_{f}(x)$ is an even function of $f$.  Observe that the missing $4\pi$ in (\ref{wef}) re-appears in (\ref{s}). For the potential (\ref{Af1})  and the Green's function (\ref{pr9})  this integration produces respectively the Li\`enard-Wiechert solution and its standard Green's function, $G(x,\tau)={1}/{r}\;\Theta(bt)\delta(r-bt)$. Retrieving the Li\`enard-Wiechert solutions could hardly be a surprise as we are summing up fields that satisfy the Lorentz gauge.
 A remarkable point is that $A_{f}$ and $G_{f}$ have no singularity in contradistinction to their continuous counterparts. This is a consequence of their distinct supports; a straight line and a lightcone, respectively. So, contrary to an old lore, the Coulomb (and also the Schwarszchild, see \cite{gr-qc/9801040}) singularity is not a consequence of a point-like source but just a reflex of the lightcone vertex \cite{BJP}. The field singularity appears with (\ref{s1s}), the integration over the lightcone.

\noindent Another remarkable point is the surging of gauge symmetry in the continuous field; even the Li\`enard-Wiechert solution, the transform of (\ref{Af1}), acquires a residual gauge freedom. In $G$, $A$ and $F$  all information implicit with each $f$ is lost with the $f$-integration. A generic solution $A$ is only indirectly linked to the state of motion of its sources; it can have many sources at once, even a continuous one or no source at all (solutions from homogeneous equation). In other words, with the $f$-integration the field $A_{f}$ becomes $A$, loses the strict constraint (\ref{dA0}) of extended causality and acquires gauge freedom. The coherence between $A_{f}$ and its source at its emission time (expressed through $f$ as a one-to-one link between a field-event and a source-event) is lost. With $A(x)$ the fibre $f$ is replaced by the lightcone and a point-charge event is linked to an infinity of field events. This is saying again that the infinity in $A$ is introduced with (\ref{s}). In this sense $A(x)$ represents more a kind of  averaging (smearing) than a union of $A_{f}\;'s$.  This explains the violation of causality in the Li\`enard-Wiechert advanced solution and the causality problems with the Lorentz-Dirac equation. Solutions with other (non-covariant) gauges just aggravates this loss.\\

This work has thrown some light on the meaning and origin of gauge fields, their symmetries and singularities.  Details and extended discussions on the physical interpretation, on the connections to classical and quantum physics and to experimental data are  left for a coming work \cite{ecwpd}. 
There are of course many other questions still to be answered; a classical scheme cannot give the complete answer. It is necessary now to distinguish what is just a consequence of the simple description adopted for the field sources from what requires a real quantum treatment. How far can we go with such a classical scheme or where a legitimate quantum input must necessarily be added?

\end{document}